\documentclass{article}

\usepackage[english]{babel}

\usepackage{tabularx}
\usepackage{float} 
\usepackage[inline]{enumitem}

\usepackage[letterpaper,top=2cm,bottom=2cm,left=3cm,right=3cm,marginparwidth=1.75cm]{geometry}

\usepackage{amsmath}
\usepackage{graphicx}
\usepackage[colorlinks=true, allcolors=blue]{hyperref}
\usepackage{cleveref}

\title{Radon Exposure Dataset}

\author{
  Dakotah Maguire\textsuperscript{1},
  Jeremy Logan\textsuperscript{1},
  Heechan Lee\textsuperscript{2}, and
  Heidi Hanson\textsuperscript{1} \\
  \\
  \textsuperscript{1}Oak Ridge National Laboratory\\
  \textsuperscript{2}Georgia Institute of Technology 
}
\date{}  

\begin{document}
\maketitle

\section{Introduction}

Exposure to elevated radon levels in the home is one of the leading causes of lung cancer in the world \cite{carmona2005radon,gonzalez2020air, gundersen1992radon}. The following study describes the creation of a comprehensive, state-level dataset designed to enable the modeling and prediction of household radon concentrations at Zip Code Tabulation Area (ZCTA) and sub-kilometer scales. Details include the data collection and processing involved in compiling physical and demographic factors for Pennsylvania and Utah. Attempting to mitigate this risk requires identifying the underlying geological causes and the populations that might be at risk. This work focuses on identifying at-risk populations throughout Pennsylvania and Utah, where radon levels are some of the highest in the country~\cite{gundersen1992radon}.

The resulting dataset harmonizes geological and demographic factors from various sources and spatial resolutions, including temperature, geochemistry, and soil characteristics. Demographic variables such as the household heating fuel used, the age of building, and the housing type provide further insight into which populations could be most susceptible in areas with potentially high radon levels.

This dataset also serves as a foundational resource for two other studies conducted by the authors.
The resolution of the data provides a novel approach to predicting potential radon exposure, and the data processing conducted for these states can be scaled up to larger spatial resolutions (e.g., the Contiguous United States [CONUS]) and allow for a broad reclassification of radon exposure potential in the United States.


\section{Census Data}

Data from the American Community Survey 5-Year (ACS5) and the Decennial Census (DEC) was collected for Pennsylvania and Utah at the block-group and ZCTA levels for the years 2000, 2013, 2015, and 2020. The variables collected are presented in \autoref{tab:variable_mapping}. Next, block-group data for each variable was converted to H3 format by using the the \texttt{area\_interpolate()} function from the PySAL Tobler library~\cite{pysal2007}. 

Differences in census table definitions across surveys require adjustments before an analysis can be conducted. 
After the 2000 DEC, the long-form survey was replaced with the American Community Survey~\cite{census_acs_diff}. The US Census Bureau and other organizations such as IPUMS provide cross walks and documentation on how to correctly modify the data to ensure the variables are as close to each other as possible for analysis~\cite{manson2024, acs_guide}. 

\clearpage
\begin{table}[H]
\centering
\caption{Census Variables Mapping and Descriptions}
\vspace{0.5em} 
\label{tab:variable_mapping}
\begin{tabularx}{\textwidth}{p{2.6cm}p{2cm}p{9.66cm}}
\hline
\textbf{DEC Code} & \textbf{ACS Code} & \textbf{Description} \\
\hline
P053001 & B19013 & Median Household Income in 1999 \\
P053 & B19013 & Median Household Income in the Past 12 Months (Inflation-Adjusted Dollars) \\
P088 & C17002 & Ratio of Income to Poverty Level in the Past 12 Months \\
H001 & B25001 & Housing Units \\
H006 & B25002 & Occupancy Status \\
H008 & B25004 & Vacancy Status \\
H030 & B25024 & Units in Structure \\
H034 & B25034 & Year Structure Built \\
H023 & B25017 & Rooms \\
H024 & B25018 & Median Number of Rooms \\
H040 & B25040 & House Heating Fuel \\
\hline
\end{tabularx}
\end{table}

\noindent Three tables in the collected data required modification. 

\begin{enumerate}[label=(\arabic*),leftmargin=*] 
\item The variables in table P088 from the 2000 DEC and table C17002 from the ACS5 needed slight adjustments due to changes in table structure made in the 2010 DEC. \textit{H008003 Total Rented or sold, not occupied} was split into \textit{Estimate Total Sold, not occupied} and \textit{Estimate Total Rented, not occupied}. For analysis, these variables were summed under the new name, \textit{B25004\_003\_005}. 

\item The \textit{H034: Year Structure Built} table in the 2000 DEC ends with the year 2000. The 1990--2000 decade was broken up into three date ranges---(1)~\textit{H034002: Built 1990~to~1994}, (2)~\textit{H034003: Built 1995~to~1998}, and (3)~\textit{H034004: Built 1999~to~March 2000}.
These three variables were summed to match the ACS5 variable \textit{B25034\_004: Estimate Total Built 1990~to~1999}.

\item Two changes were made for the variables in the P088 DEC table to reflect changes made in the 2010 DEC. \textit{P08800: Total 0.50 to 0.74} and \textit{P088004: Total 0.75 to 0.99} were combined and summed to match \textit{C17002\_003 Estimate Total 0.50 to 0.99}.
Likewise, \textit{P088007: Total 1.50 to 1.74} and \textit{P088008: Total 1.75 to 1.84} were combined to align with \textit{C17002\_006: Estimate Total 1.50 to 1.84}.  
\end{enumerate}

\subsection{Urban-Rural Classification}
 
 The DEC provides a classification system of geospatial layers at varying spatial resolutions and assigns an urban or rural designation to each component of the layer~\cite{census2020urban}. The data used was provided in a text file that contained urban block groups from the 2020 ACS5. This urban classification was appended to the other ACS5 data using the block-group geographic identifiers. Areas not classified as urban were considered rural. 

\section{Soil and Geological Data}
\begin{sloppypar}
The soil dataset was created by using the gridded National Soil Survey Geographic Database (gNATSGO) provided by the US Department of Agriculture's (USDA's) Natural Resources Conservation Service~\cite{gnatsgo2023}.
Based on the available literature~\cite{guthrie1990environmental, barros2012radon}, thirty variables that have either shown correlations or are thought to have correlations to elevated radon levels were chosen and are presented in \autoref{tab:soil_variables}.
Various soil characteristics (e.g., density, composition) can influence radon levels by affecting fluid movement and gas permeability. Additionally, soil moisture and water retention capacity can both play a role in radon levels~\cite{yang2019radon, sun2004effect}.
Other factors that may influence radon accumulation or release, such as soil surface conditions or the presence of a basement, were also included in the analysis. 
\end{sloppypar}

The ArcGIS Toolbox Soil Data Development Toolkit, designed and distributed by the US Geological Survey (USGS), and the 30-m CONUS gNATSGO grid were used to create this data.
Variable processing criteria were standardized based on the data type (continuous or categorical).
The data criteria included a soil depth of 0--200~cm, and the measure was represented as a weighted average or as a dominant condition for categorical data.
Additional processing was required to format the data to make it usable for analysis. This process is described in \textit{How To Create an On Demand Soil Property or Interpretation Grid} from gNATSGO~\cite{nrcs2022gnatsgo}.
The definitions of the variables can be found in the \textit{SSURGO Metadata Table Column Descriptions} document~\cite{ssurgo_metadata}.   

\begin{table}[H] 
\centering 
\caption{Soil Variable Names and Their Dataset Abbreviations}
\vspace{0.5em} 
\label{tab:soil_variables} 
\begin{tabular}{ll}
\hline
\textbf{Variable} & \textbf{Abbreviation} \\
\hline
Available Water Capacity WTA, 0--200~cm & AWC \\
Available Water Storage WTA, 0--200~cm & AWS \\
Available Water Supply, 0--25~cm & AWSA \\
Available Water Supply, 0--50~cm & AWSB \\
Available Water Supply, 0--150~cm & AWSC \\
Available Water Supply, 0--100~cm & AWSD \\
Percent Clay WTA, 0--200~cm & CLAY \\
Depth to Any Soil Restrictive Layer WA & DARL \\
Bulk Density, One-Third Bar WTA, 0--200~cm & DB3R \\
Drainage Class DCD & DRCL \\
Depth to a Selected Soil Restrictive Layer WTA, Abrupt textural change & DSRL \\
Dwellings With Basements DCD & DWEL \\
Dwellings Without Basements DCD & DWLO \\
Hydric Rating by Map Unit PP & HYDR \\
Hydrologic Soil Group DCD & HYSG \\
Saturated Hydraulic Conductivity (Ksat) WTA, 0--200~cm & KSAT \\
Saturated Hydraulic Conductivity (Ksat), Standard Classes WTA, 0--200~cm & KSCL \\
Linear Extensibility WTA, 0--200~cm & LEP \\
Liquid Limit WTA, 0--200~cm & LQLM \\
Organic Matter WTA, 0--200~cm & OGMT \\
Plasticity Index WTA, 0--200~cm & PLSL \\
Percent Sand WTA, 0--200~cm & SAND \\
Percent Silt WTA, 0--200~cm & SILT \\
Soil Moisture Class DCD & SMCL \\
Soil Moisture Subclass DCD & SMSC \\
Surface Texture DCD, 0--1~cm & SRFT \\
Soil Temperature Regime DCD & STMP \\
Soil Taxonomy Classification DCD & STXC \\
Water Content, 15 Bar WTA, 0--200~cm & WC15 \\
Water Content, One-Third Bar WTA, 0--200~cm & WC3R \\
\hline
\end{tabular}
\end{table}

After individual rasters were obtained for the gNATSGO variables, each raster was reprojected to EPSG:4326. Next, pixel values were extracted and reorganized into a GeoPandas DataFrame with each row containing the single pixel value along with a point geometry derived from the associated longitude and latitude of the pixel centroid.
Using those values, zonal statistics for the target hexagons were computed using \texttt{geo\_to\_h3\_aggregate} from the h3-pandas library~\cite{h3_pandas}. To avoid memory overrun, the raster data was processed by using overlapping tiles. While processing each tile, the number of pixels used to produce the hexagon value for that tile was also tracked. This allowed the selection of hexagon values for hexagons in overlapping tile areas to be from a tile in which all contributing pixels were present by simply choosing the computed hex value with the largest number of pixels.

\subsection{Lithology}
Lithological data was obtained from the USGS~\cite{anning2017}. The dataset contained 12 generalized lithological layers, which were reassigned based on the USGS state geologic map compilation for the CONUS. Various elements across different lithologic layers can contribute to increased radon levels. The data was converted to H3 format by using Python and the polyfill function from the h3-pandas library.

\subsection{Soil Geochemistry}

Increased levels of potassium, uranium, and thorium in the soil can correlate with higher concentrations of radon~\cite{usgs2007generalized}. These variables were obtained from a USGS geochemical and mineralogical survey \cite{smith2019geochemical}.  The available data was divided into three soil depths: O horizon (0--2~in.), A horizon (2--10~in.), and C horizon (30--48~in.)~\cite{nrcs_soil_profile}.  

The data was processed using the USGS-provided shapefiles and converted into H3 hexagons for ZCTA-level aggregation. GeoTIFF files that contain mineralogical data included separate bands to represent RGB values rather than a single value per element. To interpret this data, the color codes were matched with the legends provided by the USGS, which indicated categorical values. A new raster band was then generated with these categorical values and used zonal statistics (specifically, the mode statistic with the \texttt{geo\_to\_h3\_aggregate()} function from the h3pandas library) to assign representative values to each H3 level-8 hexagon. This process ensured that each hexagon in the study area had a consistent categorical representation of geochemical attributes, thereby enhancing the granularity and reliability of the radon risk predictions.

\section{Elevation Data}
Elevation data for Pennsylvania and Utah was collected from the USGS's Global Multiresolution Terrain Elevation Data 2010 (GMTED2010)~\cite{usgs2011}, which provides 30 arc-second resolution data for the United States.
After obtaining 30 arc-second resolution raster format data, the elevation data was re-gridded to H3 level-8 hexagons using the \texttt{geo\_to\_h3\_aggregate()} function from the h3pandas library \cite{h3_pandas}.
The \texttt{geo\_to\_h3\_aggregate()} function was passed raster data values, with the location of each raster pixel represented by its centroid. The function provides aggregated values for each hexagon that contains at least one raster pixel centroid by aggregating all such values according to a specified aggregation operation (mean, in this case)~\cite{h3_pandas}. To fill in missing hexagons (which occurred due to grid misalignment), the authors applied a \textit{ring smoothing} technique in which the values of missing hexagons are computed by averaging the values of adjacent hexagons~\cite{h3_pandas}. 

\section{Hydrologic Data }
  
Hydrologic variables were acquired from the USGS Hydrologic Landscape Regions dataset~\cite{usgs2023}. Four variables most relevant to hydrological influences on radon mobility were selected from the the dataset: \textit{Aquifer Permeability Class}, \textit{Minimum Elevation in Watershed}, \textit{Relief of Watershed}, and \textit{Percent Flat Land in Watershed}~\cite{tanner1994radon, sprinkel1990radon}.
The raw hydrologic data was provided as a vector dataset represented by a shapefile at a 1~km$^{2}$ scale. To align with the other datasets, the data was first read into a Geopandas GeoDataFrame, and then the \texttt{polyfill\_resample()} method from the h3-pandas library was applied to obtain values for each target H3 level-8 hexagon~\cite{h3_pandas}. 

\subsection{The GLobal HYdrogeology MaPS}
The GLobal HYdrogeology MaPS dataset provides permeability and porosity data as vector (non-gridded) data with an average polygon size of approximately 100~km$^{2}$~\cite{gleeson2014}. Soil porosity has been identified as a significant factor in radon levels~\cite{tanner1994radon, sprinkel1990radon}. Conversion from the original shapefile to H3 format was done by using Python and the polyfill function from the h3-pandas library.

\section{Meteorological Data}
The Daymet dataset consists of daily values for each variable, and each day is given as a raster layer of modeled values estimated over a 1~km$^{2}$ grid covering North America.
For the analysis, data from the CONUS was used for the following four variables during the 2008--2017 time period \cite{daymet2024}: \textit{Daily Total Precipitation}, \textit{Snow Water Equivalent}, \textit{Daily Minimum/Maximum 2-m Air Temperature}, and \textit{Vapor Pressure}.
These variables were chosen because they provide insight into the atmospheric and hydrological limitations to radon's movement. These variables can influence the airflow of a building or the emission of radon from the ground, both of which directly affect radon levels in a structure~\cite{yang2019meteo, porstendorfer2005meteo}. 

The daily Daymet grids were first aggregated into monthly averages by using NumPy, and the grid centroids were transformed from the provided Lambert Conformal Conic projection to EPSG:4326. Next, the monthly grids were re-gridded to H3 level-8 by using areal interpolation (specifically the \texttt{area\_interpolate()} function from the PySAL Tobler library), and finally a ring-smoothing technique was used to fill in missing hexagon values caused by misalignment between raster cells and the H3 hexagons~\cite{pysal2007}. 

\section{ZCTA Aggregation}
Non-census data was converted to h3 level-8 resolution. Because the radon measurements were conducted in residential buildings, hexagons that do not have residents were masked by using LandScan day and night population data. The remaining areas with population were then aggregated to ZCTA resolution. This was accomplished by using areal interpolation (specifically the \texttt{area\_interpolate()} function from the PySAL Tobler library \cite{pysal2007}). This data was then merged with the Pennsylvania ACS5 and DEC ZCTA data. 

\section{Conclusion}
This effort produced a robust, high-resolution dataset that integrates area-based measures and demographic variables that are critical for understanding household radon concentrations in Pennsylvania and Utah. By harmonizing open-source data across multiple sources and spatial scales, the resulting dataset establishes a scalable framework for assessing radon risk at sub-kilometer resolution. This resource supports more precise modeling of exposure potential and insight into the spatial variability of radon at fine geographic scales.  

\section*{Acknowledgement}
Notice: This manuscript has been authored by UT-Battelle LLC under contract DE-AC05-00OR22725 with the US 
Department of Energy (DOE). The US government retains and the publisher, by accepting the article for 
publication, acknowledges that the US government retains a nonexclusive, paid-up, irrevocable, worldwide 
license to publish or reproduce the published form of this manuscript, or allow others to do so, for US 
government purposes. DOE will provide public access to these results of federally sponsored research in 
accordance with the DOE Public Access Plan (https://www.energy.gov/doe-public-access-plan).

Oak Ridge National Laboratory's work on the LUCID: Low-dose Understanding, Cellular Insights, and Molecular Discoveries program was supported by the U.S. Department of Energy, Office of Science, Office of Biological and Environmental Research, under Contract UT-Battelle, LLC- ERKPA71

\bibliographystyle{IEEEtran} 
\bibliography{references}

\end{document}